\documentclass[a4paper,11pt]{article}
\pdfoutput=1
\usepackage{amsmath}
\usepackage{graphicx}
\usepackage{subcaption}

\usepackage{bm}
\usepackage{braket}
\usepackage{url}
\usepackage[english]{babel}
\usepackage[utf8]{inputenc}
\usepackage{pdflscape}
\usepackage{enumerate}
\usepackage{amsbsy}
\usepackage{amsmath} 
\usepackage{graphics}
\usepackage{wrapfig}
\usepackage{mathtools}

\usepackage{amsfonts}
\usepackage{pstricks}
\usepackage{color}
\usepackage{setspace}

\usepackage[normalem]{ulem}

\usepackage{accents}
\usepackage{tensor}
\usepackage [autostyle, english = american]{csquotes}
\MakeOuterQuote{"}

\usepackage{jcappub_ver} 

\newcommand{\bea}{\begin{eqnarray}} \newcommand{\eea}{\end{eqnarray}}
\newcommand{\el}{\nonumber \\}
\newcommand{\re}[1]{(\ref{#1})}

\renewcommand{\sec}[1]{section \ref{#1}}
\newcommand{\fig}[1]{figure \ref{#1}}

\renewcommand{\a}{\alpha}
\renewcommand{\b}{\beta}
\renewcommand{\c}{\gamma}
\renewcommand{\d}{\delta}

\newcommand{\ha}{\frac{1}{2}}

\newcommand{\rmd}{\mathrm{d}}

\newcommand{\ie}{i.e.\ }

\newcommand{\eq}{eq.\ }

\newcommand{\rs}{s}
\newcommand{\br}{\boldsymbol{r}}
\newcommand{\bx}{\boldsymbol{x}}
\newcommand{\by}{\boldsymbol{y}}
\newcommand{\bz}{\boldsymbol{z}}
\newcommand{\bu}{\boldsymbol{u}}
\newcommand{\bk}{\boldsymbol{k}}
\newcommand{\av}[1]{\langle{#1}\rangle}

\newcommand{\keq}{k_{\text{eq}}}
\newcommand{\kuv}{k_{\text{UV}}}
\newcommand{\kir}{k_{\text{IR}}}
\newcommand{\cP}{\mathcal{P}}

\newcommand{\cR}{\mathcal{R}}

\title{Cosmic variance and ergodicity in finite systems with correlations}

\author[a]{Dipayan Mukherjee}
\author[b]{and Syksy R\"{a}s\"{a}nen}

\affiliation[a]{Raman Research Institute, C.~V.~Raman Avenue, Sadashivanagar, Bengaluru 560080, India}

\affiliation[b]{University of Helsinki, Department of Physics and Helsinki Institute of Physics \\ P.O. Box 64, FIN-00014 University of Helsinki, Finland}

\emailAdd{dipayan@rrimail.rri.res.in}
\emailAdd{syksy.rasanen@iki.fi}

\abstract{
We consider the difference between ensemble and volume average in cosmology. It is known that for sufficiently weak long-range correlations the root mean square of the difference, which we call ergodicity bias, decays like $R^{-3/2}$ in the limit of large volume $R^3$. We calculate the condition this imposes on the power spectrum of a Gaussian random field. We quantify the bias for finite $R$, and show that the $R\to\infty$ limit is of little relevance for cosmological observations when the measured scales and correlations extend to the size of the observable universe. We consider curvature, density, and velocity perturbations. On large scales the bias is important in all three cases. For the density perturbations, which are observationally the most relevant, the relative bias first exceeds 100\% at the separation $r=177$ Mpc, and is larger than 100\% for all $r>560$ Mpc. It should be taken into account when comparing ensemble and volume averages for large-scale structure. The bias is also large for the cosmic microwave background temperature perturbations on large angular scales, but this is not relevant for observations, as their analysis does not involve volume averaging.
}

\begin{document}

\begin{flushleft}
	\hfill		 HIP-2026-13/TH \\
\end{flushleft}
 
\setcounter{tocdepth}{3}

\setcounter{secnumdepth}{3}

\maketitle

\section{Introduction}

In cosmological scenarios where primordial perturbations originate in quantum fluctuations (most notably inflation) three types of averaging appear: quantum expectation value, classical ensemble average, and volume average. When comparing theoretical predictions to observations, usually the quantum expectation value is first replaced with the average over a classical stochastic ensemble, which is then equated with the average over a spatial volume. The first step has been much studied, although not fully understood \cite{Polarski:1995jg, Lesgourgues:1996jc, Kiefer:1998qe, Kiefer:1998jk, Perez:2005gh, Kiefer:2008ku, Sudarsky:2009za, Martin:2012pea, Martin:2015qta, Martin:2019wta, Berjon:2020vdv}; we concentrate on the second. 

In cosmology, the replacement of the ensemble average with the volume average is crucial, because we can observe only one universe and hence can access only a single realisation of the ensemble. This replacement is often justified with the assumption that the system is ergodic. The concept of ergodicity comes from probability theory and statistical physics, where it has different meanings \cite{Khinchin:1949} (page 44), \cite{Gnedenko:1968} (pages 142 and 394), \cite{Adler:1981} (page 6). One of them is that time evolution samples states in such a manner that the time average over a sufficiently long period equals the ensemble average. In cosmology, volume average is used instead of time average, with the idea that all possibilities are realised in different spatial regions with the probabilities given by the probability distribution for the ensemble \cite{Gabrielli:2005} (page 34). In cosmology this equality is usually not rigorously established, nor are deviations from it quantified, and it is often regarded as a "commonsense axiom" \cite{Peacock:1999ye} (page 496) to be adopted without scrutiny. A notable exception is \cite{Weinberg:2008zzc} (page 537) (see also \cite{Pan:2010ya}), where Weinberg showed that the difference between the ensemble average and the volume average of the $n$-point function with sufficiently weak long-range correlations over a ball of radius $R$ vanishes as $R^{-3/2}$ in the $R\to\infty$ limit (compare to \cite{Khinchin:1949} (page 68)).

However, in cosmology the perturbations observed in large-scale structure and the cosmic microwave background (CMB) are correlated on all observable scales. In fact, the amplitude of the power spectrum of the primordial curvature perturbation grows with increasing wavelength, leading to long-range order called a super-homogeneous distribution, with mass fluctuations that are more ordered than for a Poisson process \cite{Gabrielli:2005} (page 80) \cite{Pietronero:2001ew, Gabrielli:2001xw}. Also, as we can observe only a finite volume, the relevance of the $R\to\infty$ limit is unclear. We extend Weinberg's calculation to include explicit correlations and calculate how rapidly the power spectrum has to fall to obtain the $R^{-3/2}$ convergence in the $R\to\infty$ limit. We also show that the $R^{-3/2}$ scaling does not guarantee suppression for realistic values of $R$. We quantify the correction due to finite volume.
 
 In \sec{sec:ergo} we calculate the ergodicity bias between ensemble and volume average. We apply it first to the comoving curvature perturbation, density contrast, and velocity perturbation in three dimensions, and then to the two-dimensional CMB temperature perturbation. In \sec{sec:conc} we summarise our findings and outline open questions. In appendix \ref{app:CMB} we derive a more numerically tractable form for the celestial ergodicity bias of the CMB.

\section{Cosmic variance and ergodicity} \label{sec:ergo}

\subsection{Ensemble and volume average} \label{sec:averages}

We mostly follow the setup and notation of \cite{Weinberg:2008zzc}. We consider a random field $\varphi(t,\bx)$; as we only study fields at equal times, we do not write down the dependence on $t$. We assume that the hypersurfaces of constant $t$ are Euclidean. We assume that the random process is homogeneous (also called stationary in statistical physics), which means that for all points $\bx_i$ and values $\bz$
\bea
  \av{ \varphi(\bx_1 + \bz) \ldots \varphi(\bx_n + \bz) } &=& \av{ \varphi(\bx_1) \ldots \varphi(\bx_n) } \ ,
\eea
where $\av{}$ stands for ensemble average. The ergodicity bias $\Delta_\varphi$ that measures the difference between volume average and ensemble average is defined by
\bea \label{Delta}
  \!\!\!\!\!\!\!\!\!\!\!\!\!\! \Delta_\varphi^2(\bm{x}_1, \ldots , \bm{x}_n, R) &\equiv& \Braket{ \left[ \int\rmd^3 z N_R(\bz) \varphi(\bx_1 + \bz) \ldots \varphi(\bx_n + \bz) - \Braket{ \varphi(\bx_1) \ldots \varphi(\bx_n) } \right]^2 } \ ,
\eea
where $N_R(\bz)$ is a window function that selects a ball of size $R$ centred on $\bz_0$ such that $\int\rmd^3 z N_R(\bz)=1$. As in \cite{Weinberg:2008zzc}, we take the window function to be Gaussian,
\bea
N_R(\bz) &=& \frac{1}{(2\pi)^{\frac{3}{2}} R^3} e^{-\ha\frac{|\bm{z}-\bm{z}_0|^2}{R^2}} \ .
\eea
The ensemble average of the volume average equals the ensemble average because of homogeneity. However, the volume average of a single realisation coincides with the ensemble average only when $\Delta_\varphi$ vanishes. We expand the square in \re{Delta}, use homogeneity, and carry out one of the volume integrals to obtain
\bea \label{Delta2}
  \Delta_\varphi^2 &=& \frac{1}{ (2\pi)^{\frac{3}{2}} R^3 } \int\rmd^3 u \, e^{-\ha\frac{u^2}{R^2}} \big[ \av{ \varphi(\bx_1 + \bu) \ldots \varphi(\bx_n + \bu) \varphi(\bx_1) \ldots \varphi(\bx_n) } \el
  && - \Braket{ \varphi(\bx_1) \ldots \varphi(\bx_n) }^2 \big] \ ,
\eea
where $u\equiv|\bu|$. At this point in \cite{Weinberg:2008zzc} it is assumed that the field is uncorrelated at distant arguments, so that in the limit $u\to\infty$ the $2n$-point expectation value decomposes into $\av{\varphi(\bx_1 + \bu) \ldots \varphi(\bx_n + \bu)} \av{\varphi(\bx_1) \ldots \varphi(\bx_n)}$, which is then by homogeneity independent of $\bu$, so the integrand vanishes. It is further assumed that this limit is approached so rapidly that the integral \re{Delta2} converges even without the exponential term, in which case the result that $\Delta_\varphi$ vanishes at least as fast as $R^{-3/2}$ follows.

Consider the two-point function $\av{\varphi(\bx + \bu) \varphi(\bx)}$. In the statistical physics of microscopic particles, $\bx$ and $\bu$ are typically time variables, so the mathematical assumption that the two-point function vanishes for $u\to\infty$ corresponds to the physical idea of "molecular chaos" \cite{Khinchin:1949} (page 67). The reasoning is that as a particle receives stochastic kicks from the environment, its properties at two times separated by a duration much longer than the kick timescale are uncorrelated. For a homogeneous Gaussian random field the vanishing of the two-point function in the limit of infinite separation guarantees ergodicity  \cite{Khinchin:1949} (page 68), \cite{Adler:1981} (page 146).

The physical situation of the cosmological three-dimensional curvature (and density and velocity) field, and the two-dimensional CMB temperature field, is different from this "molecular chaos". In the early universe, it has become correlated over all observable scales. As perturbations in the field stretch to super-Hubble scales during inflation, they become frozen (assuming they are adiabatic), so large-scale correlations are not erased. To the contrary, with increasing separation the correlations become less, not more, smeared by local evolution, as only sub-Hubble modes undergo significant evolution even after inflation. Writing the two-point function in terms of Fourier modes, we have
\bea \label{2pf}
  \!\!\!\!\!\!\!\! \av{\varphi(\bx_1) \varphi(\bx_2)} &=& \frac{1}{(2\pi)^3} \int\rmd^3 k \int\rmd^3 k' e^{i(\bk\cdot\bx_1-\bk'\cdot\bx_2)} \av{\varphi_{\bk} \varphi_{\bk'}^*} =
  \int_0^\infty\frac{\rmd k}{k} \cP_\varphi(k) j_0(k r) \ ,
\eea
where $j_0(x)=\frac{\sin x}{x}$ is the spherical Bessel function of order zero, $r\equiv|\bx_2-\bx_1|$, and we have assumed that different wavenumbers are uncorrelated and the random process is isotropic, $\av{\varphi_{\bk} \varphi_{\bk'}^*}=(2\pi)^3\d^{(3)}(\bk-\bk')\cP_\varphi(k)/(4\pi k^3)$, where $k\equiv|\bk|$. Because of homogeneity the two-point function depends only on $\br\equiv\bx_2-\bx_1$, and because of isotropy it is only a function of $r$. The integral \re{2pf} diverges at the lower end unless the power spectrum $\cP_\varphi(k)$ vanishes sufficiently fast as \mbox{$k\to0$}; any power-law with a positive power will do. (Similarly, it diverges at the upper end unless  $\cP_\varphi(k)$ vanishes sufficiently fast as \mbox{$k\to\infty$}.) In particular, the integral diverges if the power spectrum is scale-invariant or grows towards large scales, as predicted for the primordial curvature perturbation by many models of inflation and observed in the CMB and large-scale structure \cite{Planck:2018jri}. The result is then sensitive to the cutoff scale, and it is not obvious that the fall-off is sufficiently rapid to give the $R^{-3/2}$ approach to ergodicity and suppress the ergodicity bias.

To decompose the $2n$-point function in the ergodicity bias \re{Delta2} we assume, instead of cosmic molecular chaos, that the field is Gaussian (with zero mean). Odd $n$-point functions then vanish and for even $n$-point functions the square of the ergodicity bias reduces to an integral over a sum of products of two-point functions,
\bea \label{Delta2b}
  \!\!\!\!\!\!\!\!\!\!\! \Delta_\varphi^2 &=& \frac{1}{(2\pi)^{\frac{3}{2}} R^3} \int\rmd^3u \, e^{-\ha\frac{u^2}{R^2}} \Bigg[ \mathop{\sum \prod}_{\text{pairs } (i,j)} \av{\varphi(\by_i) \varphi(\by_j)}  - \Big( \mathop{\sum \prod}_{\text{pairs } (i,j)} \av{\varphi(\bx_i) \varphi(\bx_j)} \Big)^2 \Bigg] \ ,
\eea
where the sum and the product run over all pairs $i\neq j$, with $i,j=1,\ldots,2n$ in the first term and $i,j=1,\ldots,n$ in the second, and we have defined $\by_i\equiv\bx_i$ for $i=1,\ldots,n$ and $\by_i\equiv\bx_{i-n}+\bu$ for $i=n+1,\ldots,2n$. We concentrate on the two-point function, so we only need the decomposition of the four-point function,
\bea \label{dec}
  && \av{ \varphi(\bx_1+\bu) \varphi(\bx_2+\bu) \varphi(\bx_1) \varphi(\bx_2) } = \av{ \varphi(\bx_1+\bu) \varphi(\bx_2+\bu) } \av{ \varphi(\bx_1) \varphi(\bx_2) } \el
  && + \av{ \varphi(\bx_1+\bu) \varphi(\bx_1) } \av{ \varphi(\bx_2+\bu) \varphi(\bx_2) } + \av{ \varphi(\bx_1+\bu) \varphi(\bx_2) } \av{ \varphi(\bx_2+\bu) \varphi(\bx_1) } \ .
\eea
Weinberg's decomposition requirement is now satisfied if the two-point function vanishes sufficiently fast in the limit of infinite separation. Applying \re{dec}, the expression \re{Delta2}  for the square of the ergodicity bias for the two-point function reduces to
\bea \label{Delta3}
  \!\!\!\!\!\!\!\!\!\!\! \Delta_\varphi^2(r, R) &=& \frac{1}{(2\pi)^{\frac{3}{2}}} \int\rmd^3u \, e^{-\ha u^2} \big[ \av{\varphi(\bu R) \varphi(\bf 0)}^2 + \av{\varphi(\br + \bu R) \varphi(\bf 0)} \av{\varphi(\br - \bu R)\varphi(\bf 0)} \big] \ ,
\eea
where we have redefined $\bu\to\bu R$ and used homogeneity.

As noted above,  the ensemble average of the volume average equals the ensemble average because of homogeneity.  However, this is not true for the variance, and the square of the ergodicity bias \re{Delta3} is precisely the difference between the ensemble variance and the ensemble average of the volume variance. The ensemble variance of the two-point function is
\bea
  \!\!\!\!\!\!\!\!\! \sigma^2_{\varphi(E)} &\equiv& \av{ [ \varphi(\bx_1) \varphi(\bx_2) ]^2 } - \av{ \varphi(\bx_1) \varphi(\bx_2) }^2 = \av{ \varphi(\bf 0)^2 }^2 + \av{ \varphi(\br)\varphi(\bf 0) }^2 = \Delta_\varphi^2(r, 0) \ ,
\eea
where we have again assumed Gaussianity to write the four-point function in terms of the two-point function and used homogeneity. The ensemble average of the volume variance of the two-point function is
\bea
  \sigma^2_{\varphi(V)} &\equiv& \left\langle{ \int\rmd^3 z N_R(\bz) [ \varphi(\bx_1 + \bz) \varphi(\bx_2 + \bz) ]^2 - \left[ \int\rmd^3 z N_R(\bz) \varphi(\bx_1 + \bz) \varphi(\bx_2 + \bz) \right]^2 } \right\rangle \el
  &=& \Delta_\varphi^2(r, 0) - \Delta_\varphi^2(r, R) \ .
\eea
Because of homogeneity, the four-point function term is the same in the ensemble variance and in the ensemble average of the volume variance, so it is rather transparent that their difference is the square of the ergodicity bias, $\sigma^2_{\varphi(E)}-\sigma^2_{\varphi(V)}=\Delta_\varphi^2(r, R)$. We will show that $\Delta_\varphi^2(r, R)$ vanishes in the $R\to\infty$ limit, so the ensemble variance and the ensemble average of the volume variance agree. However, for any finite $R$, they are different, and the variance derived from the ensemble does not give the right error bars for the two-point function evaluated by volume average, even if the latter is averaged over many realisations.

\subsection{Three-dimensional field} \label{sec:LSS}

\subsubsection{Ergodicity bias} \label{sec:bias}

Let us calculate the second term of the ergodicity bias \re{Delta3}; the first term is then obtained from that by putting $\br=\bf 0$. Inserting the two-point function \re{2pf}, we have
\bea \label{2pfa}
  \!\!\!\!\!\!\!\!\!\!\!\!\!\!\!\!\!\!\! && \frac{1}{(2\pi)^{\frac{3}{2}}} \int\rmd^3u \, e^{-\ha u^2} \av{\varphi(\br + \bu R) \varphi(\bf 0)} \av{\varphi(\br - \bu R)\varphi(\bf 0)} \el
  \!\!\!\!\!\!\!\!\!\!\!\!\!\!\!\!\!\!\! &=& \frac{1}{(2\pi)^{\frac{3}{2}}} \int_0^\infty\frac{\rmd k}{k} \cP_\varphi(k)  \int_0^\infty\frac{\rmd k'}{k'} \cP_\varphi(k') \int\rmd^3u \, e^{-\ha u^2} j_0(k|\br+R\bu|) j_0(k'|\br-R\bu|) \el
  \!\!\!\!\!\!\!\!\!\!\!\!\!\!\!\!\!\!\! &=& \int_0^\infty\frac{\rmd k}{k} \cP_\varphi(k) \int_0^\infty\frac{\rmd k'}{k'} \cP_\varphi(k') e^{-\ha (k^2+k'^2)R^2} \sum_{\ell=0}^\infty (-1)^\ell (2\ell+1) j_\ell(kr) j_\ell(k'r) i_\ell(k k' R^2) \ ,
\eea
where $j_\ell$ and $i_\ell$ are the spherical and the modified spherical Bessel function of order $\ell$, respectively. We have used the decomposition (\eq 10.1.45 in \cite{Abramowitz:1964})
\bea \label{j0dec}
  j_0(k|\br-R\bu|) = \sum_{\ell=0}^\infty (2\ell+1) j_\ell(k r) j_\ell(k R u) P_\ell(\cos\theta) \ ,
\eea
where $P_\ell$ is the Legendre polynomial of order $\ell$ (with $\cos\theta\equiv\br\cdot\bu/(r u)$), taken into account the orthogonality of Legendre polynomials, and applied Weber's second exponential integral of two Bessel functions with a power-law times exponential kernel \cite{Offerhaus:1972}. For $r=0$, only the $\ell=0$ term contributes to the sum in \re{2pfa}, as in the $x\to0$ limit $j_\ell(x)\simeq x^\ell/(2\ell+1)!!$. Hence, the square of the ergodicity bias \re{Delta3} for the two-point function is
\bea \label{Delta4}
  \Delta_\varphi^2(r, R) &=& \int_0^\infty\frac{\rmd k}{k} \cP_\varphi(k) \int_0^\infty\frac{\rmd k'}{k'} \cP_\varphi(k') e^{-\ha (k^2+k'^2)R^2} \el
  && \times \sum_{\ell=0}^\infty ( \d_{\ell0} + 1 ) (-1)^\ell (2\ell+1) j_\ell(kr) j_\ell(k'r) i_\ell(k k' R^2) \ .
\eea
In the $R\to\infty$ limit, applying $i_\ell(x)\simeq\frac{e^x}{2x}$ for $x\gg\ha\ell(\ell+1)$ and $\lim_{x\to{\infty}} \frac{R}{\sqrt{2\pi}} e^{-\ha x^2 R^2}=\d(x)$, and using the decomposition \re{j0dec} in the other direction, \re{Delta4} reduces to
\bea \label{Deltalimit}
  \Delta_\varphi^2(r, R) &\simeq& \frac{\sqrt{\pi/2}}{R^3} \int_0^\infty\frac{\rmd k}{k^4} \cP_\varphi(k)^2 [ 1 + j_0(2kr) ] \ .
\eea
We see that the ergodicity bias of the two-point function of a Gaussian field vanishes in the limit of infinite volume as $R^{-3/2}$ provided that the momentum integral in \re{Deltalimit} converges. This requires the power spectrum to vanish fast enough as $k\to0$: a sufficient (but not necessary) condition is that $\cP_\varphi(k)$ vanishes like $k$ to a power larger than $3/2$. (Similarly, the power spectrum has to either be bounded or diverge sufficiently slowly as $k\to\infty$; scaling like a power-law with a power smaller than $3/2$ is sufficient.) This is also the condition for the integral \re{Delta2} to converge without the exponential term, as assumed by Weinberg. From \re{Deltalimit} we see that in the $R\to\infty$ limit the ergodicity bias depends only weakly on $r$. At $r=0$ the two terms in the square brackets are equal, while at large $r$ the second term is suppressed by $1/r$ and rapid oscillations, so for $r\to\infty$ the ergodicity bias drops by $1-1/\sqrt{2}\approx29\%$ compared to its value at $r=0$. From the general expression \re{Delta3} we see that the ratio $\Delta_\varphi(r, R)/\Delta_\varphi(0, R)$ approaches $1/\sqrt{2}$ as $r\to\infty$ for any value of $R$. However, in the general case, it is not obvious that $\Delta_\varphi(r, R)/\Delta_\varphi(0, R)$ cannot be larger than unity for some $r$, whereas from \re{Deltalimit} it is transparent that the maximum of $\Delta_\varphi(r, R)$ is at $r=0$.

But while the result \re{Deltalimit} shows that for infinite volume the ensemble average and the volume average are identical, it also quantifies how they are different when the volume is finite. The amplitude of the ergodicity bias $\Delta_\varphi$ depends on the scale to which $R^{-3/2}$ is compared, determined by the power spectrum.

The absolute magnitude of the bias is less relevant than its magnitude relative to the two-point function, so we define the relative ergodicity bias $d_\varphi(r, R)$ by
\bea \label{d}
  d_\varphi^2(r, R) &\equiv& \frac{ \Delta_\varphi^2(\bx_1, \bx_2)}{\av{\varphi(\bx_1) \varphi(\bx_2)}^2} \simeq \frac{\sqrt{\pi/2}}{R^3} \frac{ \int_0^\infty\frac{\rmd k}{k^4} \cP_\varphi(k)^2 [ 1 + j_0(2kr) ] }{ \left[ \int_0^\infty\frac{\rmd k}{k} \cP_\varphi(k) j_0(k r) \right]^2} \ ,
\eea
where in the second equality we have taken the limit $R\to\infty$ and inserted the two-point function \re{2pf}.

\subsubsection{Curvature perturbation}

Let us first consider a power-law spectrum, as generated by inflation for the primordial comoving curvature perturbation $\cR$. In this case the integrals in \re{d} diverge either at the lower (infrared) or the upper (ultraviolet) end or both, so we introduce cutoffs,
\bea \label{P}
  \cP_\cR (k) &=& A \left( \frac{k}{k_*} \right)^{n-1} \Theta(k-\kir) \Theta(\kuv-k) \ ,
\eea
where $\Theta$ is the step function, and $A$, $n$, $k_*$, $\kir$, and $\kuv$ are constants. The dimensionless amplitude $A$ and the pivot scale $k_*$ cancel in the relative ergodicity bias \re{d}. For numerical evaluation, we adopt the values $A=2.1\times10^{-9}$, $n=0.96$, $k_*=0.05$ Mpc$^{-1}$. They are, respectively, the observed amplitude and spectral index of the power spectrum, and the conventional pivot scale where the first two are evaluated \cite{Planck:2018jri}. We also take $\kir=10^{-4}$ Mpc$^{-1}$ and $\kuv=1$ Mpc$^{-1}$, corresponding roughly to the inverse of the current proper distance to the apparent horizon and the smallest scale in the galaxy correlation function today, respectively.

It is not clear what is a reasonable choice for $\kir$. For primordial perturbations created by cosmic inflation, the cutoff scale is determined by the beginning of inflation, and the nearly scale-invariant power spectrum is expected to extend to scales that are exponentially larger than the current apparent horizon distance. However, as modes with wavelength much larger than the apparent horizon are locally indistinguishable from a smooth background, it seems more appropriate to redefine them to be part of the local background rather than including them in the perturbations. Our choice of taking the largest possible value for the cutoff scale $\kir$, corresponding to the current apparent horizon, is conservative in the sense that, as we will see, it minimises the ergodicity bias of the three-dimensional curvature perturbation and the two-dimensional CMB perturbation. The ergodicity bias of the density contrast and the velocity perturbation turn out to be practically independent of $\kir$, as long as it is sufficiently small.

With the power spectrum \re{P}, the two-point function \re{2pf} and the ergodicity bias in the $R\to\infty$ limit \re{Deltalimit} are
\bea \label{limit1}
   \av{\cR(\bx_1) \cR(\bx_2)} &=& \frac{A}{n-1} k_*^{1-n} \left[ {}_1F_2\left( \frac{n-1}{2}; \frac{3}{2}, \frac{n+1}{2}; - \frac{1}{4} \kuv^2 r^2 \right) \kuv^{n-1} \right. \el
  && \left. - {}_1F_2 \left( \frac{n-1}{2}; \frac{3}{2}, \frac{n+1}{2}; - \frac{1}{4} \kir^2 r^2 \right) \kir^{n-1} \right] \el
  \Delta_\cR^2(r, R) &\simeq& \frac{\sqrt{\pi/2} A^2}{2n-5} k_*^{2-2n} R^{-3} \left\{ \left[ 1 + {}_1F_2\left( \frac{2n-5}{2}; \frac{3}{2}, \frac{2n-3}{2}; -\kuv^2 r^2 \right) \right] \kuv^{2n-5} \right. \el
  && \left. - \left[ 1 + {}_1F_2 \left( \frac{2n-5}{2}; \frac{3}{2}, \frac{2n-3}{2}; -\kir^2 r^2 \right) \right] \kir^{2n-5} \right\} \ ,
\eea
where $_1F_2$ is the hypergeometric function and we have assumed $n\neq1,\frac{5}{2}$. It is instructive to consider the case $r=0$, where \re{limit1} simplifies to
\bea \label{limit2}
  \av{\cR(\bx_1)^2} &=& \frac{A}{n-1} k_*^{1-n} ( \kuv^{n-1} - \kir^{n-1} ) \el
  \Delta_\cR^2(0, R) &=& \frac{\sqrt{2\pi} A^2}{2n-5} k_*^{2-2n} R^{-3} ( \kuv^{2n-5} - \kir^{2n-5} ) \ .
\eea
It is apparent that for $1-n\ll1$ and $\kuv\gg\kir$ the relative ergodicity bias in the $R\to\infty$ limit is $d_\cR(0,R)\sim(\kir R)^{-3/2}$. (For our numerical values the prefactor is $0.1$.) The $r=0$ case highlights the problem with the naive application of the $R\to\infty$ limit that the amplitude of the bias depends on the scale that combines with the $R^{-3/2}$ factor to render it dimensionless. As the integral in the ergodicity bias \re{Deltalimit} is infrared-divergent, it is dominated by the largest scale. In cosmology, the largest observed scale is also the largest scale over which it makes sense to apply the ergodic argument, $R\sim\kir^{-1}$, so there is no suppression, and the vanishing of the ergodicity bias in the mathematical $R\to\infty$ limit has little physical relevance.

\begin{figure}[t]
  \centering
  \begin{subfigure}[t]{0.49\textwidth}
    \centering
    \includegraphics[height=5.8cm]{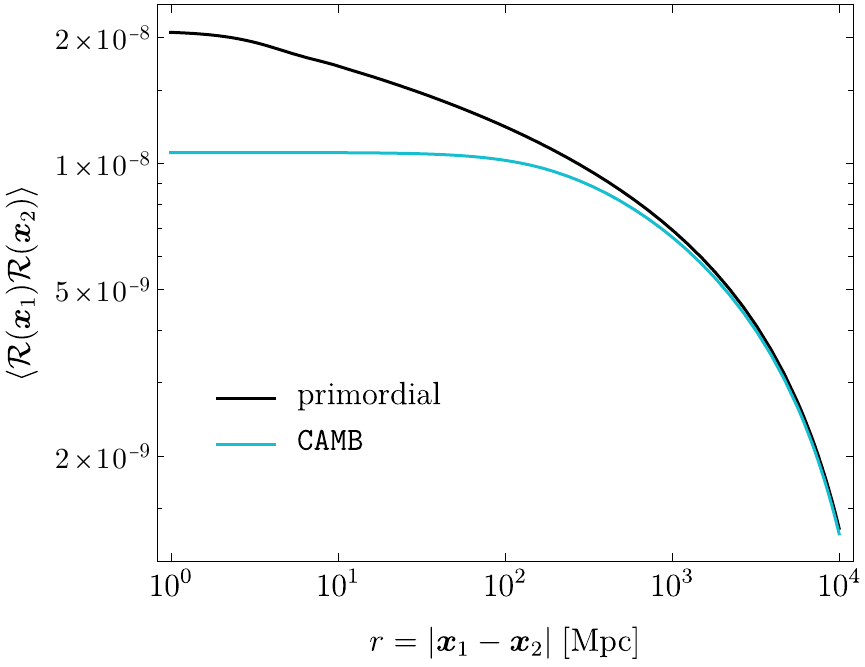}
    \caption{Two-point function.}
    \label{fig:corr_curv}
  \end{subfigure}
  \hfill
  \begin{subfigure}[t]{0.49\textwidth}
    \centering
    \includegraphics[height=5.8cm]{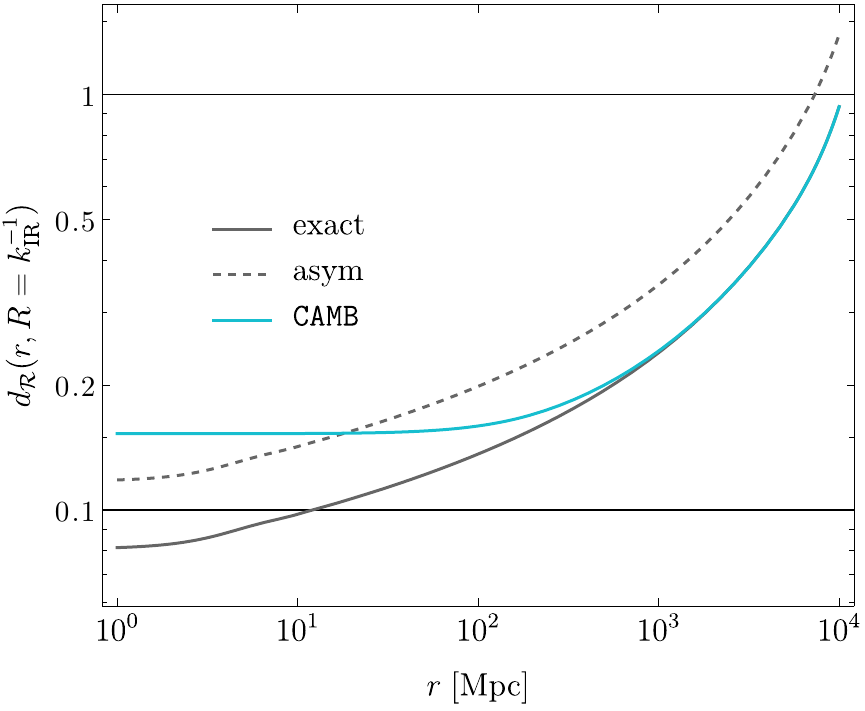}
    \caption{Relative ergodicity bias for $R=\kir^{-1}$.}
 \label{fig:normR1_curv}
  \end{subfigure}

  \vspace{0.7cm}

  \begin{subfigure}[t]{0.47\textwidth}
    \centering
    \includegraphics[height=5.4cm]{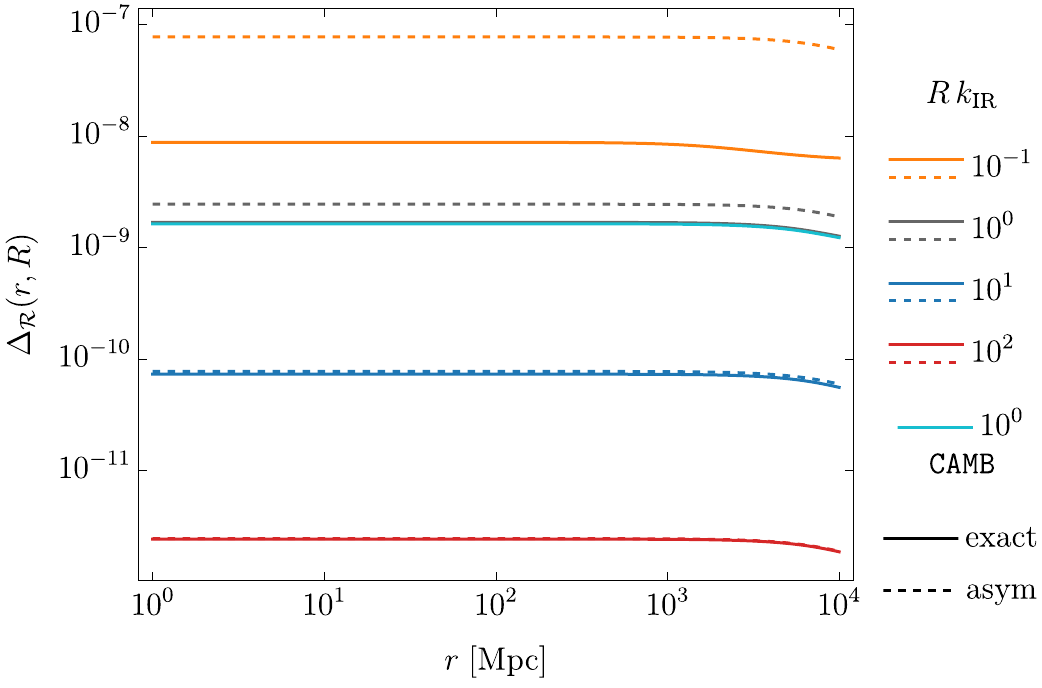}
    \caption{Ergodicity bias for different $R$.}
    \label{fig:bias_curv}
  \end{subfigure}
  \hspace{0.7cm}
  \begin{subfigure}[t]{0.47\textwidth}
    \centering
    \includegraphics[height=5.4cm]{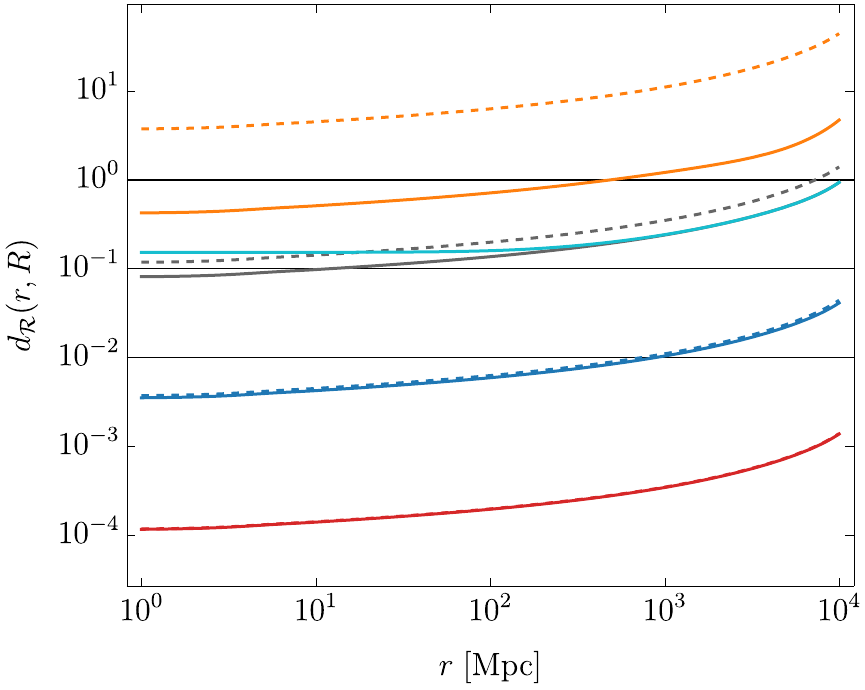}
    \caption{Relative ergodicity bias for different $R$.}
    \label{fig:norm_curv}
  \end{subfigure}
  \caption{
          \textbf{Curvature perturbation.}
          \textbf{(a)} The two-point correlation function as a function of the separation $r$. The black line is our approximation, the blue line is with the \texttt{CAMB} numerical transfer function.
          \textbf{(b)} The relative ergodicity bias $d_\cR(r, \kir^{-1})$. The line "exact" is the exact bias in our approximation, the line "asym" is the $R\to\infty$ asymptotic form \re{Deltalimit} in our approximation, and "CAMB" is with the  \texttt{CAMB} transfer function. The horizontal lines mark 10\% and 100\%. 
          \textbf{(c)} The ergodicity bias $\Delta_\cR(r, R)$ for different values of the volume-averaging scale $R$ in our approximation, and for $R=\kir^{-1}$ also for the \texttt{CAMB} transfer function.
          \textbf{(d)} The relative ergodicity bias $d_\cR(r, R)$ for the same cases as in (c). The horizontal lines mark 1\%, 10\%, and 100\% bias.
          In (b)--(d) the solid curves are the exact result and dashed curves are the $R\to\infty$ asymptotic form \re{Deltalimit}.
}
  \label{fig:curv}
\end{figure}

In \fig{fig:curv} we show the two-point correlation function, the ergodicity bias, and the relative ergodicity bias of the curvature perturbation $\cR$ as a function of the spatial separation $r$ for $\kuv^{-1}<r<\kir^{-1}$. (The values are practically unchanged from $r=0$ to $r=\kuv^{-1}$.) The two-point function in \fig{fig:curv}a has a plateau at small separation, $r\lesssim$ Mpc, and then decays logarithmically. In \fig{fig:curv}c we show the exact ergodicity bias $\Delta_{\mathcal{R}}(r, R)$ and the asymptotic $R\to\infty$ result \re{limit1} for different choices of $R$. The asymptotic result with the $R^{-3/2}$ scaling is accurate to within a factor of 2 for $R=\kir^{-1}$. (Accuracy of 10\% and 1\% is reached at $R=7\kir^{-1}$ and $R=75\kir^{-1}$, respectively, for the accurate transfer function; see below.) But the $R^{-3/2}$ scaling does not imply that the ergodicity bias is suppressed, as discussed above. As discussed in \sec{sec:bias}, $\Delta_{\mathcal{R}}(r, R)$ depends only weakly on $r$: it dips slightly as $r\approx\kir^{-1}$, and decreases by $29\%$ from $r=\kuv^{-1}$ to $r=\kir^{-1}$. The relative ergodicity bias $d_{\mathcal{R}}(r, R)$ shown in \fig{fig:curv}b for different choices of $r$ and in \fig{fig:curv}d in more detail for $R=\kir^{-1}$ is inversely proportional to the two-point function, so it grows as the two-point function decays: $d_{\mathcal{R}}(r, \kir^{-1})$ increases from $0.08$ at $r=0$ to $0.9$ at $r=\kir^{-1}$. (If the observable $r$ range were slightly larger, the relative ergodicity bias would diverge at $r=2\kir^{-1}$ where the two-point function crosses zero.) If a three-dimensional survey were to directly measure $\cR$, these values would be significant for precise comparison to theory, and conflating the volume average and the ensemble average could lead to a noticeable bias in the error analysis.

The power-law spectrum \re{P} for the curvature perturbation has been processed between primordial times and today. The most significant effect is power-law suppression and logarithmic growth of perturbations that re-entered the Hubble radius during the radiation-dominated era. The suppression approximately multiplies the perturbations by $(\keq/k)^2$ for $k\geq\keq$, where $\keq$ is the matter-radiation equality scale; we adopt the numerical value $\keq=10^{-2}$ Mpc$^{-1}$. This makes no qualitative difference for the curvature perturbation, as only the large-$k$ part of the spectrum is modified and the infrared divergence is unchanged, and even the quantitative impact is small. For the same reason the logarithmic growth has little impact. In \fig{fig:curv}a and \fig{fig:curv}b we have included the two-point function and the relative ergodicity bias $d_\cR(r,\kir^{-1})$ evaluated with an accurate transfer function at $z=0$ computed with \texttt{CAMB} that includes these effects \cite{Lewis:1999bs}, using the parameter values from Planck for the analysis with all CMB data \cite{Planck:2018jri}. For the two-point function, there is only a small offset at small $r$ compared to our approximation, and correspondingly the relative ergodicity bias is also very close. In contrast, for the density perturbation the change of evolution for modes with $k\gtrsim\keq$ is crucial and our approximation will be less accurate.

\subsubsection{Density contrast}

Observations of large-scale structure do not directly probe the curvature perturbation $\cR$, but rather the density contrast $\d$ or the velocity perturbation $v$. Dropping a prefactor of order one, in the linear regime we have $\d=(1+\frac{k^2}{3 H^2})\cR + \frac{1}{H}\dot\cR$, where dot is time derivative and $H$ is the Hubble parameter. For realistic cosmologies $H\sim\kir$, $|\dot\cR|\lesssim H|\cR|$, so as we do not consider values $k<\kir$, we approximate $\d\approx\frac{k^2}{\kir^2}\cR$. Including the $(\keq/k)^2$ suppression for modes with $k>\keq$, we have $\cP_\d=T_\d^2\cP_\cR$, where $\cP_\cR$ is the primordial power-law spectrum \re{P} and $T_\d$ is the transfer function
\bea \label{T}
T_\d(k) =
\begin{cases}
  \frac{k^2}{\kir^2} & \quad \kir \leq k \leq \keq \\
  \frac{\keq^2}{\kir^2} &\quad \keq \leq k \leq \kuv \ .
\end{cases}
\eea
This simple approximation is accurate to within an order of magnitude. Given the transfer function \re{T}, we can readily write the two-point function and the ergodicity bias analytically in terms of the hypergeometric function, analogously to \re{limit1},
\bea \label{Tlimit1}
\av{\d(\bx_1) \d(\bx_2)} &=& \frac{A}{n+3} k_*^{1-n} \kir^{-4} \left\{ \frac{1}{n+3} \left[ _1F_2\left( \frac{n+3}{2}; \frac{3}{2}, \frac{n+5}{2}; - \frac{1}{4} \keq^2 r^2 \right) \keq^{n+3} \right. \right. \el
&& \left. - _1F_2\left( \frac{n+3}{2}; \frac{3}{2}, \frac{n+5}{2}; - \frac{1}{4} \kir^2 r^2 \right) \kir^{n+3} \right] \el
&& + \frac{\keq^4}{n-1} \left[ _1F_2\left( \frac{n-1}{2}; \frac{3}{2}, \frac{n+1}{2}; - \frac{1}{4} \kuv^2 r^2 \right) \kuv^{n-1} \right. \el
&& \left. \left. - _1F_2\left( \frac{n-1}{2}; \frac{3}{2}, \frac{n+1}{2}; - \frac{1}{4} \keq^2 r^2 \right) \keq^{n-1} \right] \right\} \el
\Delta_\d^2(r, R) &\simeq& \sqrt{\pi/2} A^2 k_*^{2-2n} \kir^{-8} R^{-3} \left\{ \frac{1}{2n+3} \left[ \left( 1 + {}_1F_2\left( \frac{2n+3}{2}; \frac{3}{2}, \frac{2n+5}{2}; - \keq^2 r^2 \right) \right) \keq^{2n+3} \right. \right. \el
&& \left. - \left( 1 + {}_1F_2\left( \frac{2n+3}{2}; \frac{3}{2}, \frac{2n+5}{2}; - \kir^2 r^2 \right) \right) \kir^{2n+3} \right] \el
&& + \frac{\keq^8}{2n-5} \left[ \left( 1 + {}_1F_2\left( \frac{2n-5}{2}; \frac{3}{2}, \frac{2n-3}{2}; -  \kuv^2 r^2 \right) \right) \kuv^{2n-5} \right. \el
&& - \left. \left. \left( 1 + {}_1F_2\left( \frac{2n-5}{2}; \frac{3}{2}, \frac{2n-3}{2}; - \keq^2 r^2 \right) \right) \keq^{2n-5} \right] \right\} \ ,
\eea
where we have assumed $n\neq-3,-\frac{3}{2},1,\frac{5}{2}$. For $r=0$, the results \re{Tlimit1} simplify to
\bea \label{Tlimit2}
\!\!\!\!\!\!\!\!\!\!\!\!\!\!\!\!\! \av{\d(\bx_1)^2} &=& A k_*^{1-n} \kir^{-4} \left[ \frac{1}{n+3} ( \keq^{n+3} - \kir^{n+3} ) + \frac{\keq^4}{n-1} ( \kuv^{n-1} - \keq^{n-1} ) \right] \el
\!\!\!\!\!\!\!\!\!\!\!\!\!\!\!\!\! \Delta_\d^2(0, R) &=& \sqrt{2\pi} A^2 k_*^{2-2n} \kir^{-8} R^{-3} \left[ \frac{1}{2n+3} ( \keq^{2n+3} - \kir^{2n+3} ) + \frac{\keq^8}{2n-5} ( \kuv^{2n-5} - \keq^{2n-5} ) \right] \ .
\eea
For $1-n\ll1$ and $\kuv\gg\keq\gg\kir$ the relative ergodicity bias is $d_\d(0,R)\sim(\keq R)^{-3/2}$. For our numerical values, $d_\d(0,\kir^{-1})\approx0.3(\keq R)^{-3/2}=3\times10^{-4}$. The transfer function removes the infrared divergence and the ultraviolet part remains convergent, so the result is not sensitive to the infrared cutoff and $\keq$ becomes the dominant scale instead of $\kir$. As $R\gg\keq^{-1}$, the bias is negligible, and the asymptotic expression for the $R\to\infty$ limit is accurate.

\begin{figure*}[t]
  \centering
  \begin{subfigure}[t]{0.49\textwidth}
    \centering
    \includegraphics[height=5.8cm]{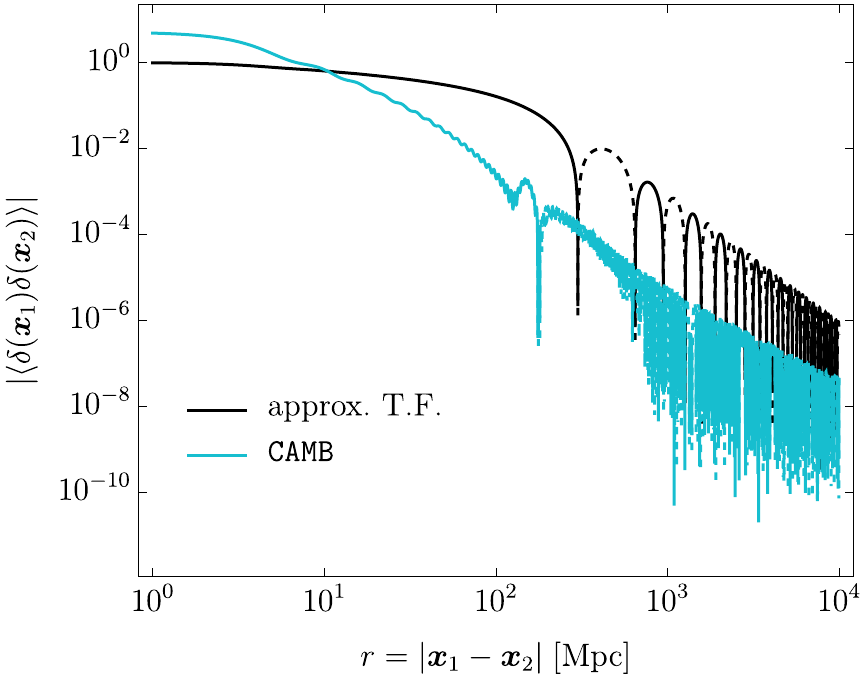}
    \caption{Two-point function.}
    \label{fig:corr_density}
  \end{subfigure}
  \hfill
  \begin{subfigure}[t]{0.49\textwidth}
    \centering
    \includegraphics[height=5.8cm]{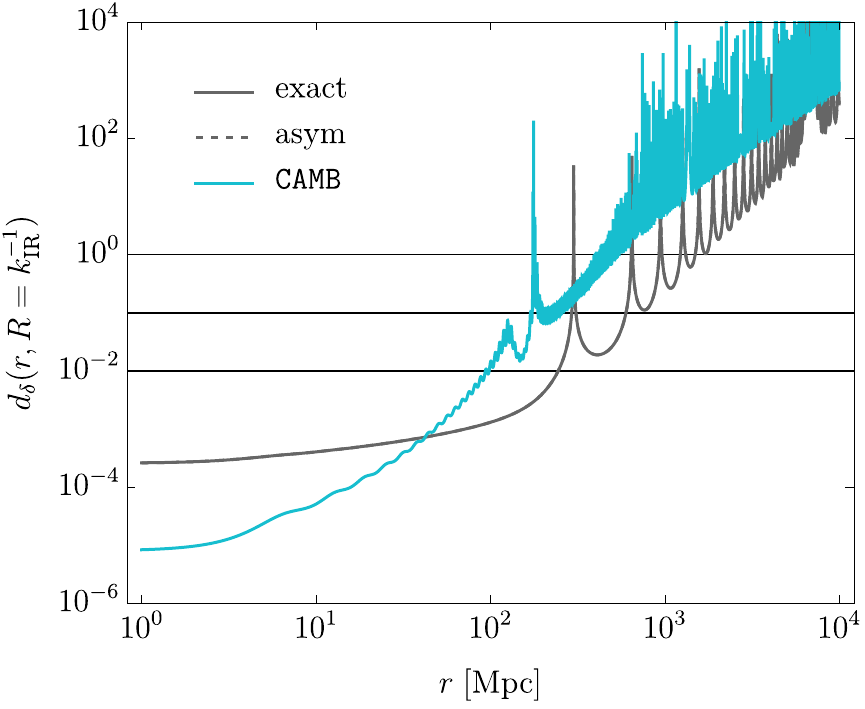}
    \caption{Relative ergodicity bias for $R=\kir^{-1}$.}
    \label{fig:normR1_density}
  \end{subfigure}

  \vspace{0.7cm}
  
  \begin{subfigure}[t]{0.47\textwidth}
    \centering
    \includegraphics[height=5.4cm]{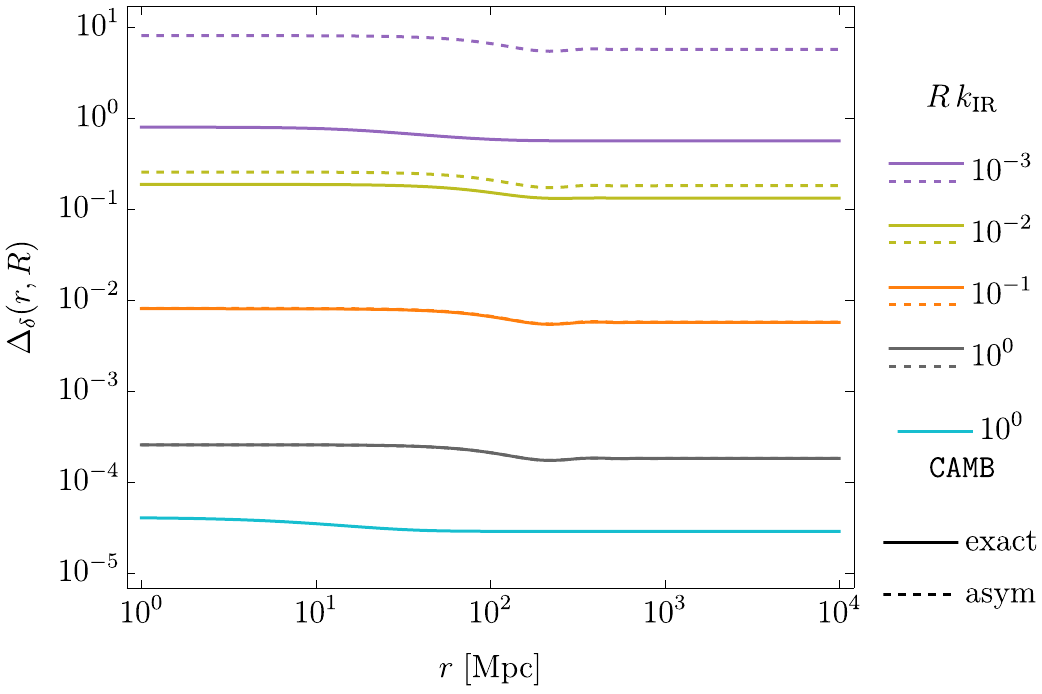}
    \caption{Ergodicity bias.}
    \label{fig:bias_density}
  \end{subfigure}
  \hspace{0.7cm}
  \begin{subfigure}[t]{0.47\textwidth}
    \centering
    \includegraphics[height=5.4cm]{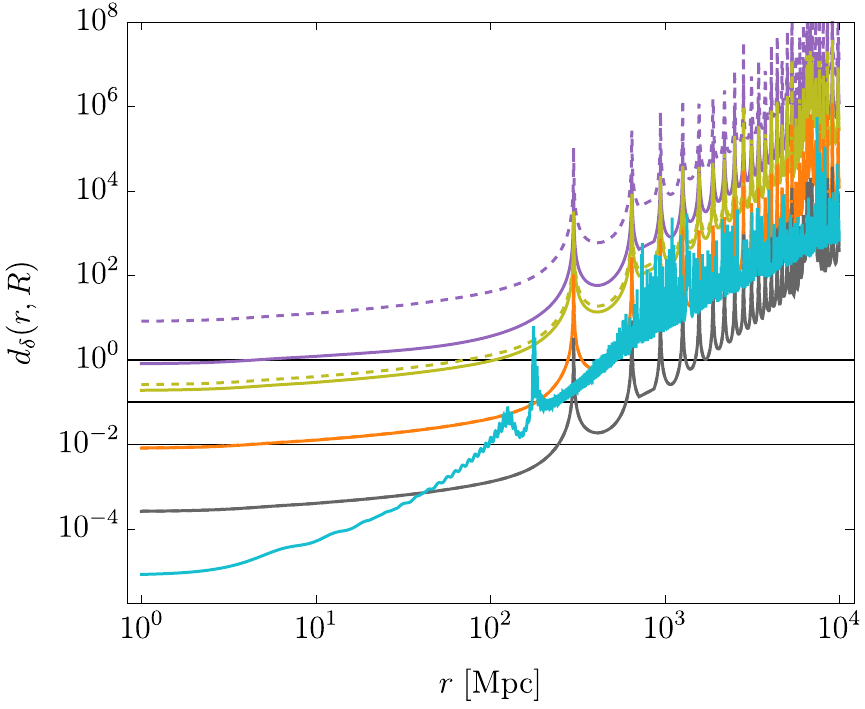}
    \caption{Relative ergodicity bias for different $R$.}
    \label{fig:norm_density}
  \end{subfigure}
  \caption{
     \textbf{Density contrast.} Same as \fig{fig:curv} but for the density contrast. In (a), dashed line indicates that the transfer function is negative.
    }
  \label{fig:delta}
\end{figure*}

In \fig{fig:delta} we show the two-point correlation function, the ergodicity bias, and the relative ergodicity bias of the density contrast $\delta$ as a function of $r$. At small $r$, the two-point correlation in \fig{fig:delta}a is constant, at $r\gtrsim$ Mpc starts to decay logarithmically, and for $r\gtrsim100$ Mpc oscillates around zero with an amplitude that decays like $r^{-2}$. From \fig{fig:delta}c we see that the ergodicity bias reaches the asymptotic $R^{-3/2}$ behaviour already for $R=10^{-1}\kir^{-1}$: the asymptotic result \re{Tlimit1} agrees with the exact expression to within $0.3\%$ everywhere. (Accuracy of 10\% and 1\% is reached already at $R=3\times10^{-3}\kir^{-1}$ and $R=1\times10^{-2}\kir^{-1}$, respectively, for the \texttt{CAMB} transfer function.) As in the curvature case, the ergodicity bias varies only little with $r$, decreasing by $29\%$ from small to large $r$. Due to the decay of the two-point function, the relative ergodicity bias $d_\d(r,\kir^{-1})$ shown in figures \ref{fig:delta}b and \ref{fig:delta}d grows with $r$. It reaches 1\% at $r=250$ Mpc and increases rapidly towards the first zero crossing of the two-point function at $r=300$ Mpc. The relative bias diverges at the zero crossings of the two-point function, of which there are many in the observable range of $r$.

Figure \ref{fig:delta}b includes the relative ergodicity bias $d_\d(r,\kir^{-1})$ evaluated with the accurate linear total matter transfer function computed with \texttt{CAMB}. At small $r$, where the bias is negligible, the realistic transfer function gives an even smaller result than our approximation, $d_\d(0,\kir^{-1})=8\times10^{-6}$ instead of $3\times10^{-4}$, but it grows faster. Neither result is reliable for $r\lesssim10$ Mpc as non-linear corrections become important. The two cross at $r=42$ Mpc. The realistic result reaches 1\% already at $r=94$ Mpc, diverges first at $r=177$ Mpc due to the zero-crossing of the two-point function, and exceeds 10\% for all $r>231$ Mpc. The approximate transfer function \re{T} captures the qualitative behaviour, but the sensitivity of the relative ergodicity bias to the two-point function when the latter is small renders the approximation inaccurate near the zero crossings. The bias is significant for large surveys, and can be neglected only at sufficiently small separation.
  
\subsubsection{Velocity perturbation}
\begin{figure*}[t]
  \centering
  \begin{subfigure}[b]{0.49\textwidth}
    \centering
    \includegraphics[height=5.8cm]{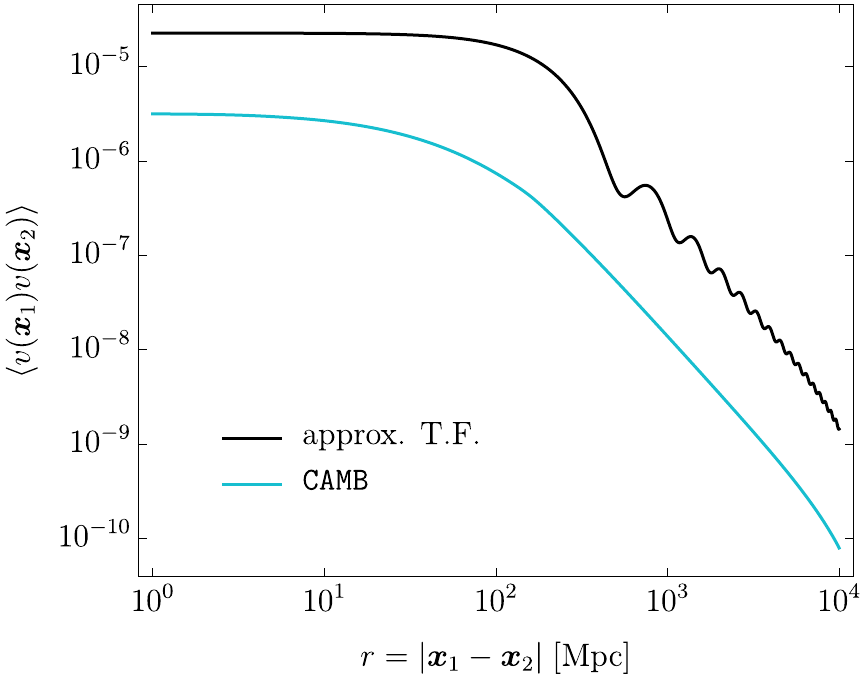}
    \caption{Two-point function.}
  \end{subfigure}
  \hfill
  \begin{subfigure}[b]{0.49\textwidth}
    \centering
    \includegraphics[height=5.8cm]{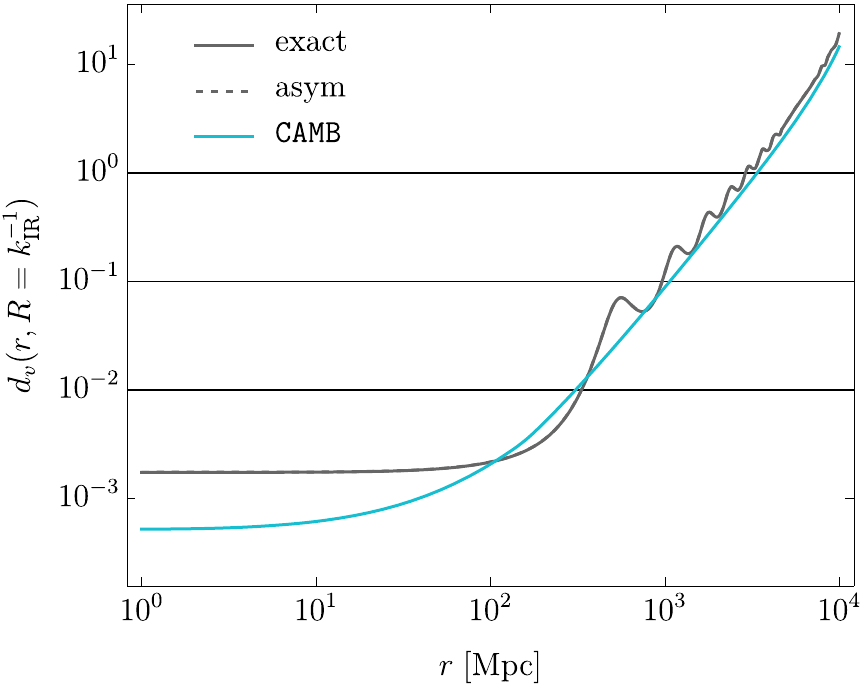}
    \caption{Relative ergodicity bias for $R = k_{\rm IR}^{-1}$.}
  \end{subfigure}

  \vspace{0.7cm}
  
  \begin{subfigure}[b]{0.47\textwidth}
    \centering
    \includegraphics[height=5.4cm]{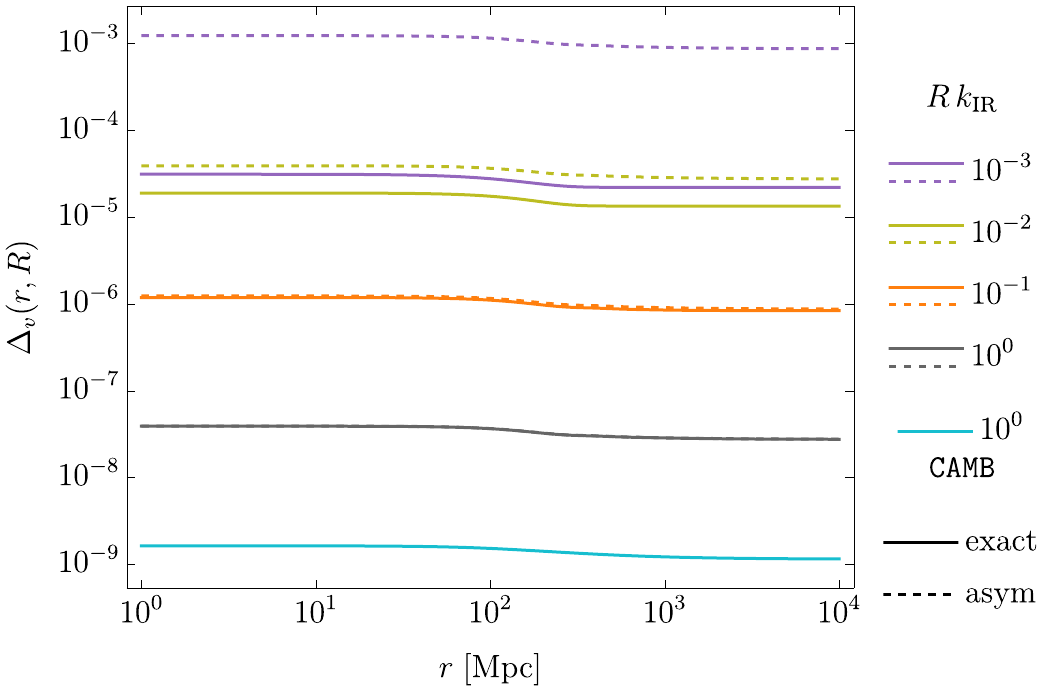}
    \caption{Ergodicity bias for different $R$.}
  \end{subfigure}
  \hspace{0.7cm}
  \begin{subfigure}[b]{0.47\textwidth}
    \centering
    \includegraphics[height=5.4cm]{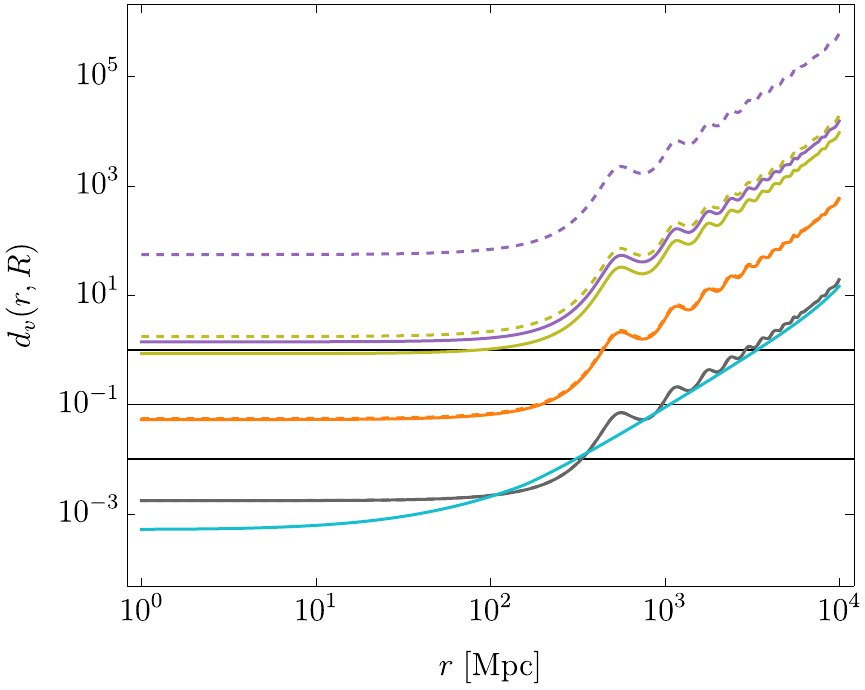}
    \caption{Relative ergodicity bias for different $R$.}
  \end{subfigure}
  \caption{
     \textbf{Velocity perturbation.} Same as \fig{fig:curv} but for the velocity perturbation.
    }
  \label{fig:v}
\end{figure*}
For the velocity perturbation $v$ we have $\cP_v=T_v^2 \cP_\cR$, where the transfer function has one factor of $k/\kir$ less than for the density contrast,
\bea \label{T2}
T_v(k) =
\begin{cases}
  \frac{k}{\kir} & \quad \kir \leq k \leq \keq \\
  \frac{\keq^2}{k \kir} &\quad \keq \leq k \leq \kuv \ .
\end{cases}
\eea
The two-point function and the ergodicity bias are readily obtained from the corresponding density contrast expressions \re{Tlimit1} by substituting $n\to n-2$ (assuming $n\neq -1, \ha, 3, \frac{9}{2}$) and multiplying  the two-point function by $(\kir/k_*)^2$ and the ergodicity bias by $(\kir/k_*)^4$. As in the case of the density contrast, the integrals are convergent in the infrared, so the results are not sensitive to the infrared cutoff. And as in the density case, the relative ergodicity bias is small at small separation, $d_v(0, R)\approx2(\keq R)^{-3/2}=2\times10^{-3}$, for $R=\kir^{-1}$.

In \fig{fig:v} we show the two-point correlation function, the ergodicity bias, and the relative ergodicity bias of the velocity perturbation $v$ as a function of $r$. The two-point function in \fig{fig:v}a is roughly constant until $r\sim10$ Mpc, then it decays, and from $r\gtrsim10^3$ Mpc it oscillates with an amplitude that decays like $r^{-2}$. The ergodicity bias is shown in \fig{fig:v}c; for $R=\kir^{-1}$, the asymptotic form is accurate to $0.6\%$. (Accuracy of 10\% and 1\% is reached at $R=0.1\kir^{-1}$ and $R=0.7\kir^{-1}$, respectively, for the \texttt{CAMB} transfer function.) The ergodicity bias as a function of $r$ behaves in the same was as in the curvature and the density cases, decreasing by $29\%$ from small to large $r$. And as in those cases, the relative ergodicity bias shown in \fig{fig:v}d grows because the two-point function decays. As in the curvature case, the first zero crossing of the two-point function is beyond the observable $r$ range, at $r=2\kir^{-1}$. The relative bias $d_v(r, \kir^{-1})$ shown in figures \ref{fig:v}b and \ref{fig:v}d first exceeds 10\% at $r=970$ Mpc, and is larger than unity for all $r>3\times10^3$ Mpc, reaching $19$ at $r=\kir^{-1}$. So as in the density case, the bias cannot be neglected at large separation.

Figure \ref{fig:v}b includes the relative ergodicity bias $d_v(r,\kir^{-1})$ calculated with the linear cold dark matter velocity transfer function at $z=0$ from \texttt{CAMB} (using the baryon velocity transfer function instead would make practically no difference). Our approximation and this realistic result agree more closely than in the density case. At small separation the realistic transfer function again gives a smaller bias, $d_v(0,\kir^{-1})=5\times10^{-4}$, about one-third of the approximate value. The two cross at $r=107$ Mpc and thereafter track each other closely. The result with the accurate transfer function reaches 1\% at 308 Mpc, 10\% at 1070 Mpc, and 100\% at 3393 Mpc. As the two-point function does not cross zero, $d_v$ does not diverge, and the simple transfer function \re{T2} gives a good approximation.
  
\subsection{CMB temperature perturbation} \label{sec:CMB}

When analysing the CMB temperature perturbation $\d T(\bx)$ (and the polarisation perturbations), the observed two-point function on the celestial sphere is directly compared to the ensemble average. The ensemble average of the difference between the square of the sky average and the square of the ensemble average is called cosmic variance, and it is central to CMB error analysis. Volume averages of theoretical quantities do not enter anywhere in the comparison between theory and experiment, unlike for the large-scale structure density contrast and the velocity perturbation. However, often cosmic variance, which is due to variation between realisations of the ensemble, is conflated with variation between spatial positions. For example, it is sometimes stated that cosmic variance is due to the observer being limited to one position in space and could be ameliorated by making observations from multiple places in the universe \cite{Kamionkowski:1997na}. It is thus interesting to see what is the error in the CMB temperature two-point function due to sampling only a finite volume, in analogy with the three-dimensional large-scale structure calculation.

The observed CMB is a pattern on the two-dimensional celestial sphere. In the ensemble average the sphere is kept fixed and the realisation of the three-dimensional field that it samples is varied. In the volume average the position of the sphere is varied across space through a fixed realisation of the three-dimensional field, or equivalently the spatial points are swept through a sphere that is kept at a fixed position. The two-point correlation function at two points separated by angle $\theta$ on the surface of a sphere with radius $\rs$ centred on $\bz$, evaluated as an average over the sphere, is
\bea \label{C}
C(\theta, \bz) &\equiv& \frac{1}{8\pi^2 \rs^4} \int\rmd^3 x_1 \int\rmd^3 x_2 \d T(\bx_1+\bz) \d T(\bx_2+\bz) \d(|\bx_1|-\rs) \d(|\bx_2|-\rs) \d( \hat\bx_1\cdot\hat\bx_2 - \cos\theta ) \el
&\equiv& \int\rmd S_1 \int\rmd S_2  \d T(\bx_1+\bz) \d T(\bx_2+\bz) \d( \hat\bx_1 \cdot \hat\bx_2 - \cos\theta ) \ ,
\eea
where we have introduced the integration measure $\rmd S_i\equiv(8\pi^2)^{-1/2} \rs^{-2} \rmd^3 x_i \d(|\bx_i|-\rs)$. The volume average of \re{C} with the window function $N_R(\bz)$ is
\bea
\bar C(\theta) &\equiv& \int\rmd^3 z N_R(\bz) C(\cos\theta, \bz) \ .
\eea
The ensemble average is instead, writing the temperature perturbation in Fourier modes and inserting the two-point function \re{2pf},
\bea \label{avC}
\av{C}(\theta) &=& \int\rmd S_1 \int\rmd S_2 \av{\d T(\bx_1) \d T(\bx_2)} \d( \hat\bx_1 \cdot \hat\bx_2 - \cos\theta )  \el
&=& \frac{1}{(2\pi)^3} \int\rmd^3 k_1 \int\rmd^3 k_2 \int\rmd S_1 \int\rmd S_2 e^{i(\bk_1\cdot\bx_1-\bk_2\cdot\bx_2)} \av{\d T_{\bk_1} \d T_{\bk_2}^*} \d( \hat\bx_1 \cdot \hat\bx_2 - \cos\theta ) \el
&=& \int\rmd^3 k \frac{\cP_{\d T}(k)}{4\pi k^3} \int\rmd S_1 \int\rmd S_2 e^{i\bk\cdot(\bx_1-\bx_2)}\d( \hat\bx_1 \cdot \hat\bx_2 - \cos\theta ) \el
&=& \sum_{\ell=0}^\infty (2\ell+1) \int_0^\infty\frac{\rmd k}{k} \cP_{\d T}(k) j_\ell(k \rs)^2 P_\ell(\cos\theta) \el
&=& \sum_{\ell=0}^\infty \frac{2\ell+1}{4\pi} C_\ell P_\ell(\cos\theta) \el
&=& \int_0^\infty\frac{\rmd k}{k} \cP_{\d T}(k) j_0( 2 k \rs \sin\tfrac{\theta}{2} ) \ .
\eea
In the fourth equality we have used the decompositions $e^{i\bk\cdot\bx}=4\pi\sum_{\ell,m} i^\ell j_\ell (kx) Y_{\ell m}(\hat\bx) Y_{\ell m}^*(\hat\bk)$ ($Y_{\ell m}$ are the spherical harmonics) and $\d(\hat\bx_1\cdot\hat\bx_2-\cos\theta)=2\pi\sum_{\ell,m} Y_{\ell m}(\hat\bx_1) Y_{\ell m}^*(\hat\bx_2) P_\ell(\cos\theta)$, and applied the orthogonality of spherical harmonics. In the last two equalities we show the result in terms of both a sum over the multipoles and an integral over the wavenumber. The next to last equality is obtained by inserting the definition $C_\ell\equiv4\pi\int_0^\infty\frac{\rmd k}{k} \cP_{\d T}(k) j_\ell(k \rs)^2$ of the angular power spectrum on a sphere of radius $\rs$. It is the standard form of the ensemble variance of the CMB two-point correlation function written in terms of multipoles. In the last equality we have applied \re{j0dec} to write the variance in terms of an integral over the three-dimensional power spectrum, which is more convenient for our purposes. As in the three-dimensional case, the ensemble average of the volume average equals the ensemble average, $\av{\bar C}=\av{C}$.

We define the celestial ergodicity bias $\tilde \Delta_{\d T}(\theta, R)$ that measures the difference between the volume average and the ensemble average analogously to the three-dimensional case,
\bea \label{DeltaTdef}
\tilde \Delta_{\d T}^2(\theta, R) &\equiv& \Big\langle \Big( \int\rmd^3 z N_R(\bz) \int\rmd S_1 \int\rmd S_2 \d( \hat\bx_1 \cdot \hat\bx_2 - \cos\theta ) \el
&& \times \big[ \d T(\bx_1 + \bz) \d T(\bx_2 + \bz) - \Braket{ \d T(\bx_1) \d T(\bx_2) } \big] \Big)^2 \Big\rangle \el
&=& \frac{2}{ (2\pi)^{\frac{3}{2}} } \int\rmd^3 u \, e^{-\ha u^2} \prod_{i=1}^4 \int\rmd S_i \d( \hat\bx_1 \cdot \hat\bx_2 - \cos\theta ) \d( \hat\bx_3 \cdot \hat\bx_4 - \cos\theta ) \el
&& \times \av{\d T(\bx_1+\bu R) \d T(\bx_3)} \av{\d T(\bx_2+\bu R) \d T(\bx_4)} \ ,
\eea
where we have expanded the square and assumed Gaussianity. The celestial ergodicity bias differs from the three-dimensional ergodicity bias \re{Delta2} via integration over spheres with the measures $\rmd S_i$ and the delta function that constrains the points to lie on the surface of a sphere of radius $\rs$, separated by angle $\theta$. Because of these factors, the two terms that remain after decomposing the four-point function and subtracting the square of the two-point function both give the same contribution, unlike in the three-dimensional ergodicity bias. 

Writing $\d T(\bx)$ in terms of Fourier modes, applying the orthogonality of spherical harmonics as before, and using the result for the integral of the product of three spherical harmonics, we obtain
\bea \label{Tbias}
\tilde \Delta_{\d T}^2(\theta, R) &=& 2 \int_0^\infty\frac{\rmd k}{k} \cP_{\d T}(k) \int_0^\infty\frac{\rmd k'}{k'} \cP_{\d T}(k') e^{-\ha (k^2+k'^2)R^2} \sum_{\ell_1,\ell_2,\ell_3} \begin{pmatrix} \ell_1 & \ell_2 & \ell_3 \\ 0 & 0 & 0 \end{pmatrix}^2 \el
&& \times (2\ell_1+1) (2\ell_2+1) (2\ell_3+1) P_{\ell_1}(\cos\theta) P_{\ell_2}(\cos\theta) \el
&& \times j_{\ell_1}(k\rs) j_{\ell_1}(k'\rs)  j_{\ell_2}(k\rs) j_{\ell_2}(k'\rs) i_{\ell_3}(k k' R^2) \ ,
\eea
where $\begin{pmatrix} \ell_1 & \ell_2 & \ell_3 \\ 0 & 0 & 0 \end{pmatrix}$ is the Wigner 3-$j$ symbol. The $k$ and $k'$ integrals are linked by $i_{\ell_3}(k k' R^2)$, with the coupling modulated by $R$. It is difficult to numerically evaluate \eqref{Tbias} because it involves a triple sum of the highly oscillatory product of four spherical Bessel functions. In appendix \ref{app:CMB} we recast the celestial ergodicity bias into a more numerically tractable form, where the sums are replaced by a single integral, which we use for generating the plots.

In the $R\to\infty$ limit, noting that $\sum_{\ell_3=0}^\infty (2\ell_3+1) \begin{pmatrix} \ell_1 & \ell_2 & \ell_3 \\ 0 & 0 & 0 \end{pmatrix}^2=1$, \re{Tbias} reduces to
\bea \label{2dbiaslimit}
\tilde \Delta_{\d T}^2(\theta, R) &\simeq& \frac{\sqrt{{2\pi}}}{R^3} \int_0^\infty\frac{\rmd k}{k^4} \cP_{\d T}(k)^2 j_0( 2 k \rs \sin\tfrac{\theta}{2} )^2 \ ,
\eea
so the bias vanishes like $R^{-3/2}$. However, as in the three-dimensional case the relevant quantity is the bias on the scales that are observed. The radius $\rs$ of the sphere is fixed, and the spatial separation is determined by the angle $\theta$. Physically $s$ is the distance to the last scattering surface, which is only 3\% smaller than the distance to the apparent horizon; for our approximate evaluation, we take $\rs=\kir^{-1}$.

Unlike the density contrast and the velocity perturbation, the temperature perturbation is not enhanced on small scales relative to the curvature perturbation. Instead, it is affected by acoustic oscillations and damping. As a simple approximation, we neglect those effects, and simply consider the power-law spectrum inherited from the primordial curvature perturbation. This is accurate for large angles, $\theta\gtrsim2\pi/30\approx0.2$, where $\d T=-\frac{1}{5}\cR T_0$; here $T_0=2.725$ K is the CMB mean temperature. The integrals in the relative ergodicity bias are thus infrared-divergent, as in the case of the three-dimensional curvature perturbation, so we do not expect the ergodicity bias to be small when $R$ is of the order of the infrared cutoff. For $0.2\lesssim\theta\leq\pi$, the separation $r=2\rs\sin(\theta/2)$ varies from $\approx0.2\rs$ to $2\rs$.

We define the relative celestial ergodicity bias $\tilde d_{\d T}(\theta, R)$ by
\bea \label{d2}
\tilde d_{\d T}^2(\theta, R) &\equiv& \frac{ \tilde \Delta_{\d T}^2(\theta, R) }{ \av{C}(\theta)^2 } \simeq \frac{\sqrt{2\pi}}{R^3} \frac{ \int_0^\infty\frac{\rmd k}{k^4} \cP_{\d T}(k)^2 j_0( 2 k \rs \sin\tfrac{\theta}{2} )^2 }{ \left[ \int_0^\infty\frac{\rmd k}{k} \cP_{\d T}(k) j_0( 2 k \rs \sin\tfrac{\theta}{2} ) \right]^2 } \ .
\eea
This is similar to the three-dimensional result \re{d} with separation $r=2\rs\sin(\theta/2)$: the only difference is that there is $j_0^2$ instead of $1+j_0$ in the numerator.

\begin{figure*}[tbp]
  \centering
  \begin{subfigure}[b]{0.49\textwidth}
    \centering
    \includegraphics[height=5.8cm]{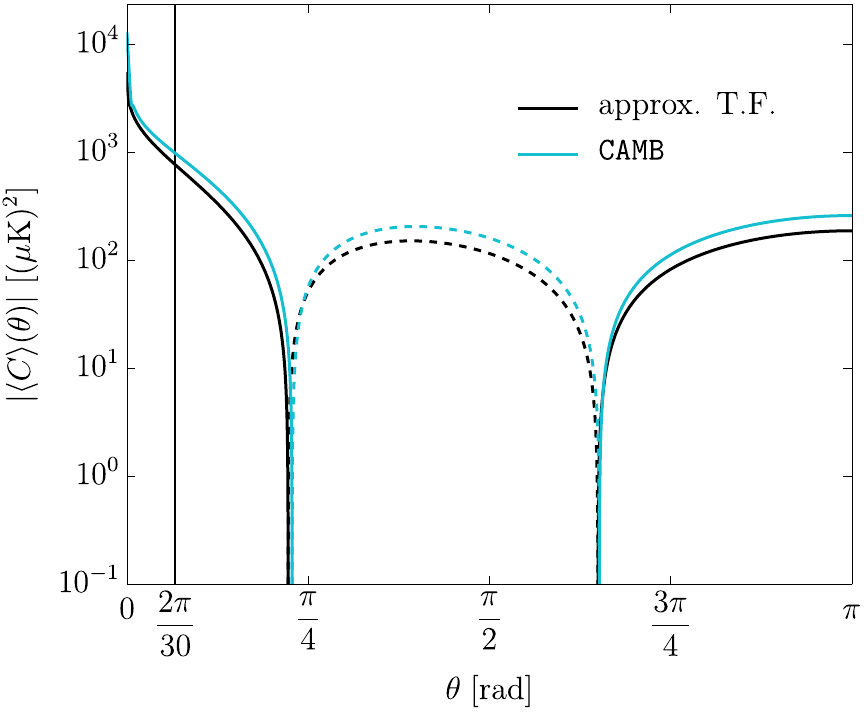}
    \caption{Angular two-point function.}
  \end{subfigure}
  \hfill
  \begin{subfigure}[b]{0.49\textwidth}
    \centering
    \includegraphics[height=5.8cm]{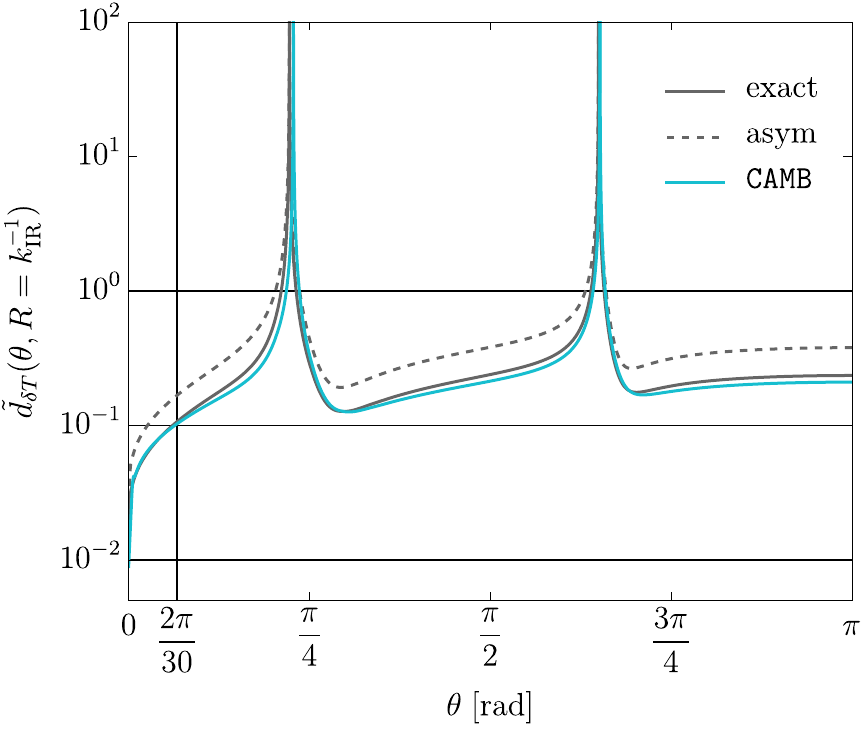}
    \caption{Relative celestial bias at $R = k_{\rm IR}^{-1}$.}
  \end{subfigure}

  \vspace{0.7cm}
  
  \begin{subfigure}[b]{0.49\textwidth}
    \centering
    \includegraphics[height=5.5cm]{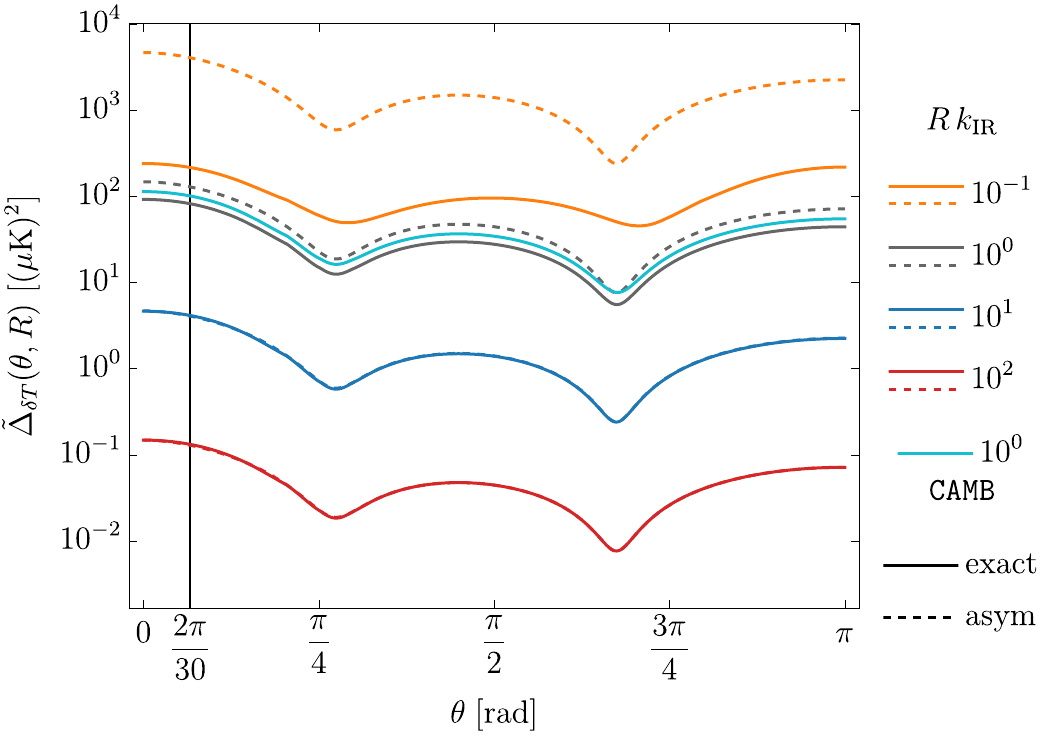}
    \caption{Celestial ergodicity bias.}
  \end{subfigure}
  \hfill
  \begin{subfigure}[b]{0.49\textwidth}
    \centering
    \includegraphics[height=5.5cm]{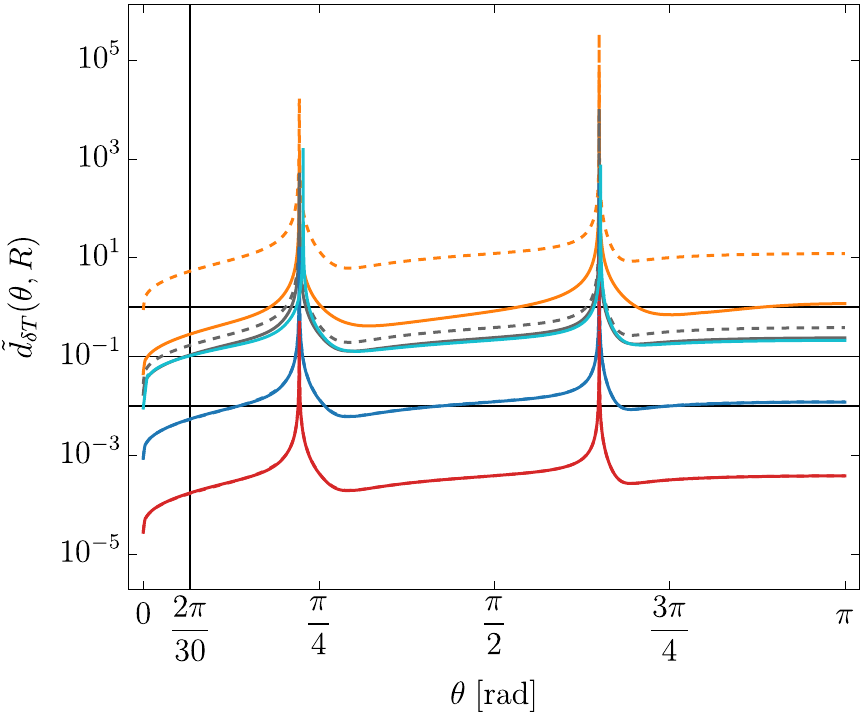}
    \caption{Relative celestial ergodicity bias.}
  \end{subfigure}

  \caption{
       \textbf{CMB temperature perturbation.} Similar to \fig{fig:curv}, but for the CMB temperature perturbation as a function of the angular separation $\theta$. In (a) and (c) the scale is in $(\mu$K$)^2$. In (a), dashed line indicates that the transfer function is negative.
       }
  \label{fig:cmb}
\end{figure*}

We show results evaluated both with our large-angle and power-law approximation, and with the full CMB transfer function $\Delta_\ell(k)$ computed with \texttt{CAMB} (not to be confused with the celestial ergodicity bias $\tilde\Delta_{\d T}(\theta,R)$). The form \re{C} for the two-point correlation function, which samples values of $\delta T$ on a sphere, is strictly speaking only valid when the path-dependent impact of propagation from the last scattering surface, such as the integrated Sachs--Wolfe effect, can be neglected. For an accurate description, these effects have to be included. We map the spectrum given by \texttt{CAMB} to our description as follows. In \texttt{CAMB}, the function $\Delta_\ell(k)$ determines the angular power spectrum via $C_\ell=4\pi\int_0^\infty\frac{\rmd k}{k}\cP_\cR(k)\Delta_\ell(k)^2$.  Comparing this with our expression $C_\ell=4\pi\int_0^\infty\frac{\rmd k}{k}\cP_{\d T}(k) j_\ell(k\rs)^2$, we have
\bea \label{PdT}
  \cP_{\d T}(k) &=& \cP_\cR(k) \frac{    \sum_\ell (2\ell+1) \Delta_\ell(k)^2 } { \sum_{\ell'} (2{\ell'}+1)    j_{\ell'}(k\rs)^2 } \ .
\eea
Analytically, the sum goes from $\ell=0$ to $\ell=\infty$, in which case $\sum_\ell(2\ell+1)j_\ell(x)^2=1$. However, numerically it is cut off at a finite value (we sum up to $\ell=2500$), so we keep the denominator. Here, we use the accurate value $\rs=1.3873\times 10^4$ Mpc for the distance to the last scattering surface rather than the approximation $\rs=\kir^{-1}$.

In \fig{fig:cmb} we show the angular two-point correlation function, the celestial ergodicity bias, and the relative celestial ergodicity bias of the CMB temperature perturbation $\d T$ as a function of the angular separation $\theta$. In the angular correlation function (and the bias) we have subtracted the monopole and dipole, as is conventional, \ie dropped the $\ell=0,1$ terms in \re{avC}. The vertical line at $\theta=\frac{2\pi}{30}$ marks the end of the validity of our large-angle approximation; the \texttt{CAMB} result is valid on all scales. Unlike in the density and velocity plots, all scales are in the linear regime. Our approximation is rather good, despite the fact that the two-point function crosses zero twice, as the locations of the zero-crossings are well reproduced. For $R=\kir^{-1}$, the bias $\tilde \Delta_{\d T}(\theta, R)$ has not yet converged to the asymptotic $R^{-3/2}$ behaviour, there is still a factor of 3 difference. (Accuracy of 10\% and 1\% is reached at $R=4\kir^{-1}$ and $R=17\kir^{-1}$, respectively, for the \texttt{CAMB} transfer function.) Unlike in the three-dimensional cases, the celestial ergodicity bias $\tilde \Delta_{\d T}(\theta, R)$ has significant evolution, but like there, the main factor that determines the change of the relative bias is the decrease of the two-point function, given that it crosses zero. Except near the zero-crossings, the relative celestial ergodicity bias at $R=\kir^{-1}$ is between 10\% and 100\% for all large angles, $\theta>\frac{2\pi}{30}$.

As noted above, volume averages do not enter into the usual CMB analysis, so the celestial ergodicity bias has no impact on the interpretation of CMB data measured from our position only, and is mainly of theoretical interest.

\section{Conclusions} \label{sec:conc}

We have quantified the ergodicity bias, defined as the root mean square difference between ensemble average and volume average, in cosmology. We have shown that the ergodicity bias of the two-point function vanishes for a Gaussian random field in the limit of large volume $R^3$ like $R^{-3/2}$ if the power spectrum vanishes sufficiently fast in the limit of small wavenumber $k$; a power-law with a power larger than $3/2$ is sufficient. However, the $R\to\infty$ limit is of little relevance for cosmological observations. In the statistical physics of systems with microscopic constituents, ergodicity is understood in terms of time average as opposed to volume average, and systems can be measured over times much larger than the microscopic timescale that characterises their correlations. In contrast, in cosmology we can observe only a finite volume whose scale is not orders of magnitude larger than the separations considered, and correlations extend over all observables scales, so the $R^{-3/2}$ scaling does not imply suppression of the ergodicity bias. 

When analysing large-scale structure data, $n$-point functions are calculated from the volume average of observed quantities. They are compared to theoretical predictions calculated from the ensemble, as well as to volume averages found from simulations that have been initialised by distributing the matter according to the ensemble probability distribution. One simulation corresponds to a single realisation of the ensemble. Ergodicity bias quantifies the difference between the two-point function calculated from the ensemble and from a single simulation (or from observations, which are also a single realisation). The ensemble average of the volume average is equal to the ensemble average, so averaging the two-point function over simulations gives an unbiased estimate of the ensemble average. However, the ensemble variance and volume variance are not equal, and their difference is precisely the square of the ergodicity bias.

We have considered the three-dimensional curvature perturbation, density contrast, and velocity perturbation, as well as the two-dimensional CMB temperature anisotropy. For the curvature perturbation, we have taken a power-law primordial spectrum with infrared and ultraviolet cutoffs. To get analytical insight, we have used an approximation for the linear transfer function that is correct to within an order of magnitude. We have also used a transfer function computed with \texttt{CAMB} for accurate numerical results. (All numbers quoted below are with the \texttt{CAMB} transfer function.) Our approximation  is reasonably accurate, especially on large scales where the ergodicity bias relative to the two-point function is significant. The notable exception is the density case, where the zero-crossing locations of the two-point function are not well reproduced, leading to large errors. Near the zero-crossings, it might be more informative to compare the ergodicity bias to the magnitude of the observational errors in the two-point function rather than the theoretical two-point function.

For all the three-dimensional quantities, the ergodicity bias is almost independent of the separation $r$, decreasing only by 29\% from small to large $r$. But because the two-point function decays with $r$ as correlations get weaker, the relative ergodicity bias (ergodicity bias divided by the two-point function) grows with $r$. For the curvature perturbation, both the two-point function and the ergodicity bias are infrared-divergent, \ie sensitive to the largest scale in the power spectrum. The asymptotic $R^{-3/2}$ scaling is reached at 10\% accuracy at $R=7\kir^{-1}$, $R=3\times10^{-3}\kir^{-1}$, and $R=0.1\kir^{-1}$ in the curvature, density, and velocity case, respectively, and at 1\% accuracy for values of $R$ one order of magnitude larger. (Here $\kir^{-1}$ is the size of the observable universe as well as the largest correlation scale.) However, as noted above, the $R^{-3/2}$ scaling does not imply that the ergodicity bias is small. 

At small $r$, the relative ergodicity bias (with $R=\kir^{-1}$) is $0.2$, $8\times10^{-6}$, and $5\times10^{-4}$ for the curvature, density contrast, and velocity, respectively. However, for $r\lesssim10$ Mpc, our linear theory result is unreliable as non-linear corrections are important. For the curvature, the relative bias rises to 0.9 for the largest separation $r=\kir^{-1}$. For the density contrast it reaches 10\% at $r=174$ Mpc and 100\% at $r=177$ Mpc. For the velocity perturbation, the corresponding numbers are $r=1070$ Mpc and $r=3393$ Mpc. For the curvature perturbation (unlike  for the density contrast and the velocity perturbation) the two-point correlation is not directly measured, so its ergodicity bias is not relevant for present observations, although its transverse gradient can be extracted from tomographic lensing observations and cross-correlations with other perturbations. For the density and the velocity field of large-scale structure, the bias is significant and should be taken into account when comparing ensemble and volume averages.

For the CMB, the data analysis is different. There the observed two-point function on the celestial sphere is directly compared to the ensemble average, and volume averaging does not enter. The usual cosmic variance error bars for the CMB quantify the error due to being able to observe only a single realisation, not the error due to being able to observe the CMB from only a single position, unlike sometimes claimed. We have calculated the celestial ergodicity bias that gives the difference between the two. It is significant for large angular separations, but this is not directly relevant for the comparison of theory and observations.

We have assumed that the fields are Gaussian when decomposing the four-point function in terms of the two-point function. It is straightforward to extend the analysis to general $n$-point functions and cross-correlations of perturbations, and include non-Gaussianity. Statistical homogeneity is a key assumption in our analysis, and relaxing it would introduce additional contributions to the ergodicity bias \cite{Ragavendra:2024qpj}. We have considered Euclidean spatial slices. Cosmological observations are consistent with small average spatial curvature \cite{Planck:2018jri}, but observations probe the past lightcone instead of a spatial hypersurface. The difference is important when survey depth is comparable to the Hubble radius, and it would be interesting to extend the calculation of the ergodicity bias to the lightcone.

\section*{Acknowledgements}

DM thanks Shiv K.~Sethi and H.~V.~Ragavendra for valuable discussions during the initial phase of this work. SR thanks Till Sawala, Giorgio Torrieri, and Jussi Väliviita for useful discussions. DM thanks Raman Research Institute for support through postdoctoral fellowship. M365 Copilot based on GPT-5 has been used to help in calculating integrals; all calculations have been checked by hand or with Mathematica.

\appendix

\section{Integral form of the celestial ergodicity bias} \label{app:CMB}

The celestial ergodicity bias \re{Tbias} for the CMB temperature perturbation involves a triple sum over multipoles and a double integral over wavenumbers of the product of four spherical Bessel functions. This expression can be highly oscillatory, making it difficult to evaluate accurately, so we recast the celestial ergodicity bias into a more manageable integral form.

Starting from \re{DeltaTdef}, expanding $\delta T (\bm{x})$ in Fourier modes and the delta functions in spherical harmonics and Legendre polynomials, and performing the spatial integrals and the integration over $\bm{u}$, we obtain
\bea \label{Tbiasint}
\tilde{\Delta}_{\delta T}^{2}(\theta, R)
&=&
2 \int \frac{{\rm d} ^3k}{4\pi k^3}
\mathcal{P}_{\delta T}(k)
\int \frac{{\rm d} ^3k'}{4 \pi k'^3}
\mathcal{P}_{\delta T}(k') e^{-\frac{1}{2}(\bm{k}-\bm{k}')^2 R^2}
W(k, k', \a, \b)^2 \el
&=&  \int_0^\infty \frac{{\rm d}k}{k} \mathcal{P}_{\delta T}(k) \int_0^\infty \frac{{\rm d}k'}{k'} \mathcal{P}_{\delta T}(k') \int_{-1}^{1} {\rm d}\a \; e^{-\frac{1}{2}(k^2+k'^2-2kk'\a)R^2} W(k, k', \a, \b)^2 \ , \el
\eea
where $\a \equiv \hat{\bm{k}} \cdot \hat{\bm{k}}'$, $\b\equiv\cos\theta$, and
\bea \label{W}
W(k, k', \a, \b) &\equiv& \sum_\ell (2 \ell + 1) P_{\ell} (\a) P_\ell(\b) j_\ell (k \rs) j_\ell (k' \rs) \el
&=& \frac{1}{2\pi} \int_0^{2\pi} \rmd\psi \sum_\ell (2 \ell + 1) P_\ell(\cos\c) j_\ell (k \rs) j_\ell (k' \rs) \el
&=&\frac{1}{2\pi} \int_0^{2\pi} \rmd\psi j_0( \rs \sqrt{ k^2 + k'^2 - 2 k k' \cos\c }) \ ,
\eea
where $\cos\gamma \equiv \a\b + \sqrt{1-\a^2}\sqrt{1-\b^2} \cos\psi$. In the second equality of \re{W} we have used the identity
\begin{equation}
  P_\ell(\a) P_\ell(\beta) = \frac{1}{2\pi} \int_0^{2\pi} \rmd\psi P_\ell(\cos\gamma) \ ,
\end{equation}
which follows from the Legendre polynomial addition theorem, and in the last equality of \re{W} we have applied the identity \re{j0dec}. The integral \re{W} can be calculated analytically. However, as the form \re{Tbiasint}, together with \re{W}, contains no sums, only integrals, it is already sufficiently stable, and we use it for numerical evaluation of the celestial ergodicity bias.

\bibliographystyle{JHEP}
\bibliography{ergo}

\end{document}